\documentclass{article}
\usepackage{amsmath,amsthm}
\usepackage{epsfig}
\newtheorem{theorem}{Theorem}
\newtheorem{proposition}{Proposition}[section]
\newtheorem{corollary}[proposition]{Corollary}
\newtheorem{lemma}[proposition]{Lemma}
\theoremstyle{definition}
\newtheorem{definition}[proposition]{Definition}
\newtheorem{example}[proposition]{Example}

\title{Mining All Non-Derivable Frequent Itemsets}

\author{Toon Calders\footnote{Research Assistant of the Fund for Scientific Research - Flanders (FWO-Vlaanderen).} \\
University of Antwerp, Belgium
\and Bart Goethals \\ University of Limburg, Belgium} 

\date{}

\begin{document}
\maketitle

\begin{abstract}
Recent studies on frequent itemset mining algorithms resulted in significant
performance improvements. However, if the minimal support threshold is set too low,
or the data is highly correlated, the number of frequent itemsets
itself can be prohibitively large. To overcome this problem, recently
several proposals have been made to construct a concise representation
of the frequent itemsets, instead of mining all frequent itemsets.
The main goal of this paper is to identify redundancies in the set of
all frequent itemsets and to exploit these redundancies in order to reduce
the result of a mining operation. We present deduction rules to derive tight
bounds on the support of candidate itemsets. We show how the deduction rules
allow for constructing a minimal representation for all frequent itemsets.
We also present connections between our proposal and
recent proposals for concise representations and we give the results of
experiments on real-life datasets that show the
effectiveness of the deduction rules. In fact, the experiments even show that
in many cases, first mining the concise representation, and then creating the
frequent itemsets from this representation outperforms existing frequent set mining
algorithms.
\end{abstract}

\section{Introduction}
The frequent itemset mining problem \cite{AGRAWAL93} is by now well known.
We are given a set of items $\cal I$ and a database $\cal D$ of subsets
of $\cal I$, together with a unique identifier. The elements of $\cal D$
are called transactions. An \emph{itemset} $I \subseteq {\cal I}$ is some
set of items; its \emph{support} in $\cal D$, denoted by ${\it support}(I,{\cal D})$,
is defined as the number of transactions in $\cal D$ that contain all items of $I$;
and an itemset is called $s$-\emph{frequent} in $\cal D$ if its support in $\cal D$
exceeds $s$. $\cal D$ and $s$ are omitted when they are clear from the context.
The goal is now, given a minimal support threshold and a database, to find all
frequent itemsets.

The search space of this problem, all subsets of $\cal I$, is
clearly huge. Instead of generating and counting the supports of all these itemsets
at once, several solutions have been proposed to perform a more directed search
through all patterns. However, this directed search enforces several scans through
the database, which brings up another great cost, because these databases tend to
be very large, and hence they do not fit into main memory.

The standard Apriori
algorithm \cite{AGRAWAL94} for solving this problem is based on the
\emph{monotonicity property}: all supersets of an infrequent itemset must be
infrequent. Hence, if an itemset is infrequent, then all of its supersets can be
\emph{pruned} from the search-space. An itemset is thus considered potentially
frequent, also called a \emph{candidate} itemset, only if all its subsets are
already known to be frequent. In every step of the algorithm, all candidate itemsets
are generated and their supports are then counted by performing a complete scan of
the transaction database. This is repeated until no new candidate itemsets can be
generated.

Recent studies on frequent itemset mining algorithms resulted in significant
performance improvements. In the early days, the size of the database and the
generation of a reasonable amount of frequent itemsets were considered the most
costly aspects of frequent itemset mining,
and most energy went into minimizing the number of scans through the
database. However, if the minimal support threshold is set too low,
or the data is highly correlated, the number of frequent itemsets
itself can be prohibitively large. To overcome this problem, recently
several proposals have been made to construct a concise representation
of the frequent itemsets, instead of mining all frequent
itemsets~\cite{PASQUIER99,BASTIDE2000,BOULICAUT2000,BOULICAUT2000b,PEI2000b,ZAKI99,BYKOWSKI2001,Kryszkiewicz2001}.

\paragraph{Our contributions} The main goal of this paper is to present several new
methods to identify redundancies in the set of all frequent itemsets and to exploit these
redundancies, resulting in a concise representation of all frequent itemsets and
significant performance improvements of a mining operation.
\begin{enumerate}

\item
We present a complete set of \emph{deduction rules\/} to derive \emph{tight\/} intervals on
the support of candidate itemsets.

\item
We show how the deduction rules can be used to construct a \emph{minimal representation\/}
of all frequent itemsets, consisting of all frequent itemsets of which the exact support can not
be derived, and present an algorithm that efficiently does so.

\item Also based on these deduction rules, we present an efficient
method to find the exact support of all frequent itemsets, that
are not in this concise representation, \emph{without scanning the
database\/}.

\item
We present \emph{connections\/} between our proposal and recent
proposals for concise representations, such as \emph{free
sets\/}~\cite{BOULICAUT2000}, \emph{disjunction-free
sets\/}~\cite{BYKOWSKI2001}, and \emph{closed
sets\/}~\cite{PASQUIER99}. We also show that known tricks to
improve performance of frequent itemset mining algorithms, such as
used in MAXMINER~\cite{BAYARDO98} and PASCAL~\cite{BASTIDE2000},
can be described in our framework.

\item We present several \emph{experiments\/} on real-life datasets that show the
effectiveness of the deduction rules.
\end{enumerate}

The outline of the paper is as follows. In Section 2 we introduce
the deduction rules. Section 3 describes how we can use the rules
to reduce the set of frequent itemsets. In Section 4 we give an
algorithm to efficiently find this reduced set, and in Section 5
we evaluate the algorithm empirically. Related work is discussed
in depth in Section 6.

\section{Deduction Rules}
In all that follows, $\cal I$ is the set of all items and $\cal D$
is the transaction database.

We will now describe sound and complete rules for deducing tight
bounds on the support of an itemset $I \subseteq {\cal I}$, if the
supports of all its subsets are given.  In order to do this, we
will not consider itemsets that are no subset of $I$, and we can
assume that all items in $\cal D$ are elements of $I$.  Indeed,
``projecting away'' the other items in a transaction database does
not change the supports of the subsets of $I$.
\begin{definition} {\bf ($I$-Projection)} Let $I \subseteq {\cal I}$ be an itemset.
\begin{itemize}
\item The \emph{$I$-projection of a transaction $T$}, denoted
$\pi_I T$, is defined as $\pi_I T := \{i \mid i \in T \cap I \}.$
\item The \emph{$I$-projection of a transaction database $\cal
D$}, denoted $\pi_I {\cal D}$, consist of all $I$-projected
transactions from $\cal D$.
\end{itemize}
\end{definition}
\begin{lemma} Let $I,J$ be itemsets, such that $I \subseteq J \subseteq {\cal I}$.
For every transaction database $\cal D$, the following holds:
$$ {\it support}(I,{\cal D}) = {\it support}(I,\pi_{J}{\cal D}). $$
\end{lemma}
Before we introduce the deduction rules, we introduce fractions
and covers.
\begin{definition} {\bf ($I$-Fraction)} Let $I,J$ be itemsets, such that $I \subseteq
J \subseteq {\cal I}$, the \emph{$I$-fraction} of $\pi_J{\cal D}$, denoted by
$f_I^J({\cal D})$ equals the number of transactions in $\pi_J{\cal D}$ that exactly
consist of the set $I$.
\end{definition}
If $\cal D$ is clear from the context, we will write $f_I^J$, and
if $J={\cal I}$, we will write $f_I$. The support of an itemset
$I$ is then $\sum_{I \subseteq I' \subseteq {\cal I}} f_{I'}.$
\begin{definition} {\bf (Cover)} Let $I \subseteq {\cal I}$ be an itemset.
The \emph{cover} of $I$ in $\cal D$, denoted by ${\it Cover}(I,{\cal D})$,
consists of all transactions in $\cal D$ that contain $I$.
\end{definition}
Again, we will write ${\it Cover}(I)$ if $\cal D$ is clear from the context.

Let $I,J \subseteq {\cal I}$ be itemsets, and $J = I \cup \{A_1,
\ldots, A_n\}$. Notice that ${\it Cover}(J)=\bigcap_{i=1}^n {\it
Cover}(I\cup \{A_i\})$, and that $|\bigcup_{i=1}^n{\it
Cover}(I\cup\{A_i\})|=|{\it Cover}(I)|-f_I^J$. From the well-known
\emph{inclusion-exclusion principle}~\cite[p.181]{KNUTHIE97} 
we learn
\begin{multline*} |{\it Cover}(I)|-f_I^J = \sum_{1 \leq i \leq n} |{\it Cover}(I \cup
\{A_i\})| \\ 
 - \sum_{1 \leq i < j \leq n} |{\it Cover}(I \cup \{A_i,A_j\})| + \cdots - (-1)^{n} |{\it Cover}(J)|,
\end{multline*}
and since ${\it support}(I \cup \{A_{i_1},\ldots,A_{i_\ell}\}) = |{\it Cover}(I \cup \{A_{i_1},
\ldots,A_{i_\ell}\})|$, we obtain
\begin{multline*} (-1)^{|J-I|} {\it support}(J) - f_I^J = {\it
support}(I) - \sum_{1 \leq i \leq n} {\it support}(I \cup \{A_i\})
\\ {} + \sum_{1 \leq i < j \leq n} {\it support}(I \cup \{A_i,A_j\})
+ \cdots + (-1)^{|J-I|-1} \sum_{1 \leq i \leq n} {\it support}(J -
\{A_i\} )
\end{multline*}
From now on, we will denote the sum on the right-hand side of this last equation
by ${\sigma}(I,J)$.

Since $f_I^J$ is always positive, we obtain the following theorem.
\begin{theorem}
For all itemsets $I,J\subseteq {\cal I}$, ${\sigma}(I,J)$ is a
lower (upper) bound on ${\it support\/}(J)$ if $|J-I|$ is even
(odd). The difference $|{\it support\/}(J)-{\sigma}(I,J)|$ is
given by $f_I^J$. \label{the:rules}
\end{theorem}
We will refer to the rule involving $\sigma(I,J)$ as ${\cal
R}_J(I)$ and omit $J$ when clear from the context.

If for each subset $I\subset J$, the support ${\it
support}(I,{\cal D})=s_I$ is given, then the rules ${\cal R}_J(.)$
allow for calculating lower and upper bounds on the support of
$J$. Let $l$ denote the greatest lower bound we can derive with
these rules, and $u$ the smallest upper bound we can derive. Since
the rules are sound, the support of $J$ must be in the interval
$[l,u]$. In \cite{CALDERSEDBT2002}, we show also that these bounds
on the support of $J$ are \emph{tight}; i.e., for every smaller
interval $[l',u']\subset [l,u]$, we can find a database $\cal D'$
such that for each subset $I$ of $J$, ${\it support}(I,{\cal
D'})=s_I$, but the support of $J$ is not within $[l',u']$.
\begin{theorem}
For all itemsets $I,J\subseteq {\cal I}$, the rules $\{{\cal
R}_J(I) \mid I\subseteq J\}$ are sound and complete for deducing
bounds on the support of $J$ based on the supports of all subsets
of $J$.
\end{theorem}
The proof of the completeness relies on the fact that for all
$I\subseteq J$, we have ${\it support}(I,{\cal D})=\sum_{I \subseteq I' \subseteq {\cal I}} f_{I'}.$
We can consider the linear program consisting of all these equalities, together with the conditions
$f_I\geq 0$ for all fractions $f_I$. The existence of a database $\cal D'$ that satisfies the given
supports is equivalent to the existence of a solution to this linear program in the $f_I$'s
and ${\it support}(J,\cal D')$. From this equivalence, tightness of the bounds can be proved.
For the details of the proof we refer to \cite{CALDERSEDBT2002}.
\begin{figure}[tbh]
\begin{equation*}
\left\{\begin{array}{lcll}
s_{ABCD}&\geq&s_{ABC}+s_{ABD}+s_{ACD}+s_{BCD}-s_{AB}-s_{AC}-s_{AD}&{\cal R}_{\{\}}
\\&&-s_{BC}-s_{BD}-s_{CD}+s_A+s_B+s_C+s_D-s_{\{\}}
\\s_{ABCD}&\leq&s_{A}-s_{AB}-s_{AC}-s_{AD}+s_{ABC}+s_{ABD}+s_{ACD}  &{\cal R}_A
\\s_{ABCD}&\leq&s_{B}-s_{AB}-s_{BC}-s_{BD}+s_{ABC}+s_{ABD}+s_{BCD} &{\cal R}_B
\\s_{ABCD}&\leq&s_{C}-s_{AC}-s_{BC}-s_{CD}+s_{ABC}+s_{ACD}+s_{BCD}&{\cal R}_C
\\s_{ABCD}&\leq&s_{D}-s_{AD}-s_{BD}-s_{CD}+s_{ABD}+s_{ACD}+s_{BCD}&{\cal R}_{D}
\\s_{ABCD}&\geq&s_{ABC}+s_{ABD}-s_{AB}&{\cal R}_{AB}
\\s_{ABCD}&\geq&s_{ABC}+s_{ACD}-s_{AC}&{\cal R}_{AC}
\\s_{ABCD}&\geq&s_{ABD}+s_{ACD}-s_{AD}&{\cal R}_{AD}
\\s_{ABCD}&\geq&s_{ABC}+s_{BCD}-s_{BC}&{\cal R}_{BC}
\\s_{ABCD}&\geq&s_{ABD}+s_{BCD}-s_{BD}&{\cal R}_{BD}
\\s_{ABCD}&\geq&s_{ACD}+s_{BCD}-s_{CD}&{\cal R}_{CD}
\\s_{ABCD}&\leq&s_{ABC}&{\cal R}_{ABC}
\\s_{ABCD}&\leq&s_{ABD}&{\cal R}_{ABD}
\\s_{ABCD}&\leq&s_{ACD}&{\cal R}_{ACD}
\\s_{ABCD}&\leq&s_{BCD}&{\cal R}_{BCD}
\\s_{ABCD}&\geq&0&{\cal R}_{ABCD}
\end{array}\right.
\end{equation*}
\caption{Tight bounds on $s_{ABCD}$. $s_I$ denotes ${\it support}(I)$}\label{fig:boundsABCD}
\end{figure}

\begin{example}
Consider the following transaction database.
$${\cal D}=\begin{array}{|l|}
\hline
  A,B,C
\\A,C,D
\\A,B,D
\\C,D
\\B,C,D
\\A,D
\\B,D
\\B,C,D
\\B,C,D
\\A,B,C,D
\\\hline
\end{array}\qquad
\begin{array}{lccclccclcc}
  s_{A}  &=&5,&   \quad   &s_{B}  &=&7,&   \quad   &s_{C}  &=&7,
\\s_{D}  &=&9,&           &s_{AB} &=&3,&   \quad   &s_{AC} &=&3,
\\s_{AD} &=&4,&           &s_{BC} &=&5,&           &s_{BD} &=&6,
\\s_{CD} &=&6,&           &s_{ABC}&=&2,&           &s_{ABD}&=&2,
\\s_{ACD}&=&2,&           &s_{BCD}&=&4.&&&&
\end{array}$$
Figure~\ref{fig:boundsABCD} gives the rules to determine tight
bounds on the support of $ABCD$. Using these deduction rules,
we derive the following bounds on $s_{ABCD}$ {\em without
counting in the database\/}.
\begin{tabbing}
Lower bound:\qquad\=$s_{ABCD}\geq 1$ \qquad\=(Rule ${\cal R}(AC)$)
\\Upper bound: \>$s_{ABCD}\leq 1$ \>(Rule ${\cal R}(A)$)
\end{tabbing}
Therefore, we can conclude, without having to rescan the database, that the
support of $ABCD$ in $\cal D$ is exactly $1$, while a standard monotonicity check
would yield an upper bound of $2$.
\end{example}

\section{Non-Derivable Itemsets as a Concise Representation}
Based on the deduction rules, it is possible to generate a summary
of the set of frequent itemsets. Indeed, suppose that the
deduction rules allow for deducing the support of a frequent
itemset $I$ {\em exactly\/}, based on the supports of its subsets.
Then there is no need to explicitly count the support of $I$
requiring a complete database scan; if we need the support of $I$,
we can always simply derive it using the deduction rules. Such a
set $I$, of which we can perfectly derive the support, will be
called a {\em Derivable Itemset\/} (DI), all other itemsets are
called {\em Non-Derivable Itemsets\/} (NDIs). We will show in this
section that the set of frequent NDIs allows for computing the
supports of all other frequent itemsets, and as such, forms a {\em
concise representation\/} \cite{MANNILA96} of the frequent
itemsets. To prove this result, we first need to show that when a
set $I$ is non-derivable, then also all its subsets are
non-derivable. For each set $I$, let $l_I$ ($u_I$) denote the
lower (upper) bound we can derive using the deduction rules.

\begin{lemma} {\bf (Monotonicity)}
Let $I\subseteq {\cal I}$ be an itemset, and $i\in {\cal I}-I$ an item. Then
$2|u_{I\cup\{i\}}-l_{I\cup\{i\}}|\leq 2\min(|{\it support}(I)-l_I|, |{\it support}(I)-u_i|)\leq |u_I-l_I|$.
In particular, if $I$ is a DI, then also $I\cup\{i\}$ is a DI.
\label{lem:lemma1}
\end{lemma}
\begin{proof}
The proof is based on the fact that $f_{J}^I=f_{J}^{I\cup\{i\}}+f_{J\cup\{I\}}^{I\cup\{i\}}$.
From Theorem~\ref{the:rules} we know that $f_{J}^I$ is the difference between the bound
calculated by ${\cal R}_I(J)$ and the real support of $I$. Let now $J$ be such that the rule
${\cal R}_I(J)$ calculates the bound that is closest to the support of $I$. Then, the width
of the interval $[l_I,u_I]$ is at least $2 f_J^I$. Furthermore, ${\cal R}_{I\cup\{i\}}(J)$
and ${\cal R}_{I\cup\{i\}}(J\cup\{i\})$ are a lower and an upper bound on the support of
$I\cup\{i\}$ (if $|I\cup\{i\}-(J\cup\{i\})|$ is odd, then $|I\cup\{i\}-J|$ is even and vice versa),
and these bounds on $I\cup\{i\}$ differ respectively $f_{J}^{I\cup\{i\}}$ and $f_{J\cup\{I\}}^{I\cup\{i\}}$ from
the real support of $I\cup\{i\}$. When we combine all these observations, we get: $u_{I\cup\{i\}}-l_{I\cup\{i\}}\leq f_{J}^{I\cup\{i\}}+
f_{J\cup\{I\}}^{I\cup\{i\}}=f_{J}^I\leq \frac{1}{2} (u_I-l_I)$.
\end{proof}

This lemma gives us the following valuable insights.
\begin{corollary}\label{cor:cor1}
The width of the intervals exponentially shrinks with the size of the itemsets.
\end{corollary}
This remarkable fact is a strong indication that the number of large NDIs will
be very small. This reasoning will be supported by the results of the experiments.
\begin{corollary}\label{cor:cor2}
If $I$ is a NDI, but it turns out that ${\cal R}_I(J)$ equals the
support of $I$, then all supersets $I\cup\{i\}$ of $I$ will be a
DI, with rules ${\cal R}_{I\cup\{i\}}(J)$ and ${\cal
R}_{I\cup\{i\}}(J\cup\{i\})$.
\end{corollary}
We will use this observation to
avoid checking all possible rules for $I\cup\{i\}$. This avoidance
can be done in the following way: whenever we calculate bounds on
the support of an itemset $I$, we remember the lower and upper
bound $l_I,u_I$. If $I$ is a NDI; i.e., $l_I\neq u_I$, then we
will have to count its support. After we counted the support,
 the tests ${\it support}(I)=l_I$ and ${\it support}(I)=u_I$ are performed. If one of these two
equalities obtains, we know that all supersets of $I$ are
derivable, without having to calculate the bounds.
\begin{corollary}
If we know that $I$ is a DI, and that rule ${\cal R}_I(J)$ gives
the exact support of $I$, then ${\cal R}_{I\cup\{i\}}(J\cup\{i\})$
gives the exact support for $I\cup\{i\}$.
\end{corollary}\label{cor:cor3}
Suppose that we
want to build the entire set of frequent itemsets starting from
the concise representation. We can then use this observation to
improve the performance of deducing all supports. Suppose we need
to deduce the support of a set $I$, and of a superset $J$ of $I$;
instead of trying all rules to find the exact support for $J$, we
know in advance, because we already evaluated $I$, which rule to
choose. Hence, for any itemset which is known to be a DI, we
only have to compute a single deduction rule to know its exact support.

From Lemma~\ref{lem:lemma1}, we easily obtain the following theorem, saying that the set of NDIs is
a concise representation. We omit the proof due to space limitations.
\begin{theorem}
For every database $\cal D$, and every support threshold $s$, let $\mathrm{NDI}({\cal D},s)$ be the following set:
$$\mathrm{NDI}({\cal D},s):=\{(I,{\it support}(I,{\cal D}))\mid l_I\neq u_I\}.$$
$\mathrm{NDI}({\cal D},s)$ is a concise representation for the
frequent itemsets, and for each itemset $J$ not in
$\mathrm{NDI}({\cal D},s)$, we can decide whether $J$ is frequent,
and if $J$ is frequent, we can exactly derive its support from the
information in $\mathrm{NDI}({\cal D},s)$.
\end{theorem}

\section{The NDI-Algorithm}
Based on the results in the previous section, we propose a
level-wise algorithm to find all frequent NDIs. Since derivability
is monotone, we can prune an itemset if it is derivable. This
gives the NDI-algorithm as shown below. The correctness of the
algorithm follows from the results in
Lemma~\ref{lem:lemma1}. 
\begin{tabbing}
\qquad\=\qquad\=\qquad\=\qquad\=\qquad\=\qquad\=\qquad\=\qquad\=\qquad\kill
\\\>NDI(${\cal D}$,$s$)
\\\>\>$i:=1$; $\mathrm{NDI}:=\{\}$; $C_1:=\{\{i\}\mid i\in {\cal I}\};$
\\\>\>\textbf{for all} $I$ in $C_1$ \textbf{do} $I.l:=0; I.u:=|{\cal D}|;$
\\\>\>\textbf{while} $C_i$ not empty \textbf{do}
\\\>\>\>Count the supports of all candidates in $C_i$ in one pass over $\cal D$;
\\\>\>\>$F_i := \{I\in C_i\mid {\it support}(I,{\cal D})\geq s\};$.
\\\>\>\>$\mathrm{NDI}:=\mathrm{NDI}\cup F_i;$
\\\>\>\>$Gen:=\{\};$
\\\>\>\>\textbf{for all} $I\in F_i$ \textbf{do}
\\\>\>\>\>\textbf{if} ${\it support}(I)\neq I.l$ and ${\it support}(I)\neq I.u$ \textbf{then}
\\\>\>\>\>\>$Gen:=Gen\cup\{I\};$
\\\>\>\>$PreC_{i+1}:=AprioriGenerate(Gen);$
\\\>\>\>$C_{i+1}:=\{\};$
\\\>\>\>\textbf{for all} $J\in PreC_{i+1}$ \textbf{do}
\\\>\>\>\>Compute bounds $[l,u]$ on support of $J$;
\\\>\>\>\>\textbf{if} $l \neq u$ \textbf{then} $J.l:=l; J.u:=u; C_{i+1}:=C_{i+1}\cup\{J\};$
\\\>\>\>$i:=i+1$
\\\>\>\textbf{end while}
\\\>\>\textbf{return} NDI
\end{tabbing}
Since evaluating all rules can be very cumbersome, in the
experiments we show what the effect is of only using a couple of
rules. We will say that we use rules {\em up to depth $k$\/} if we
only evaluate the rules ${\cal R}_J(I)$ for $|I-J|\leq k$. The
experiments show that in most cases, the gain of evaluating rules
up to depth $k$ instead of up to depth $k-1$ typically quickly
decreases if $k$ increases. Therefore, we can conclude that in
practice most pruning is done by the rules of limited depth.

\section{Experiments}

For our experiments, we implemented an optimized version of the
Apriori algorithm and the NDI algorithm described in the previous
section.  We performed our experiments on several real-life
datasets with different characteristics, among which a dataset
obtained from a Belgian retail market, which is a sparse dataset
of $41\,337$ transaction over $13\,103$ items.  The second dataset
was the BMS-Webview-1 dataset donated by Z. Zheng et
al.~\cite{realworld}, containing $59\,602$ transactions over $497$
items. The third dataset is the dense census-dataset as available
in the UCI KDD repository~\cite{UCIKDD}, which we transformed into
a transaction database by creating a different item for every
attribute-value pair, resulting in $32\,562$ transactions over
$22\,072$ items. The results on all these datasets were very
similar and we will therefore only describe the results for the
latter dataset.

\begin{figure}
\begin{center}
\epsfig{file=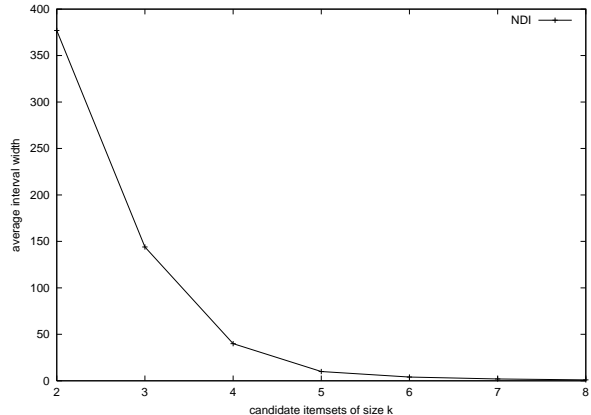,width=8cm} \caption{Average interval-width of
candidate itemsets.}\label{fig:widths}
\end{center}
\end{figure}

Figure~\ref{fig:widths} shows the average width of the intervals
computed for all candidate itemsets of size $k$.
Naturally, the interval-width of the singleton
candidate itemsets is $32\,562$, and is not shown in the figure.
In the second pass of the NDI-algorithm, all candidate itemsets of
size $2$ are generated and their intervals deduced.
As can be seen, the average interval size of most candidate itemsets
of size 2 is $377$.
From then on, the interval sizes decrease exponentially as was predicted by
Corollary~\ref{cor:cor1}.

\begin{figure}
\begin{center}
\epsfig{file=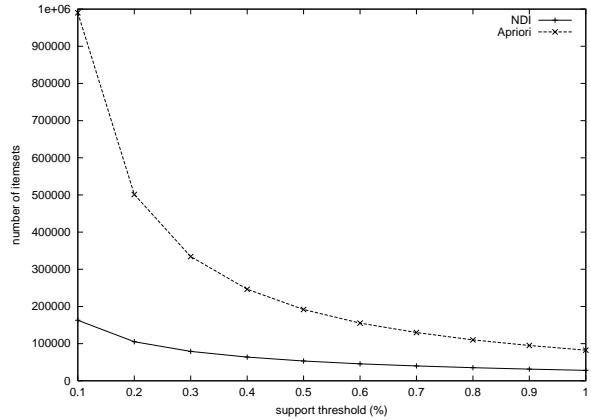,width=8cm} \caption{Size of concise
representation.}\label{fig:crsupp}
\end{center}
\end{figure}

Figure~\ref{fig:crsupp} shows the size of the concise
representation of all NDIs compared to the total number of
frequent patterns as generated by Apriori, for varying minimal
support thresholds. If this threshold was set to $0.1\%$, there
exist $990\,097$ frequent patterns of which only $162\,821$ are
non-derivable.  Again this shows the theoretical results obtained
in the previous sections.

\begin{figure}
\begin{center}
\epsfig{file=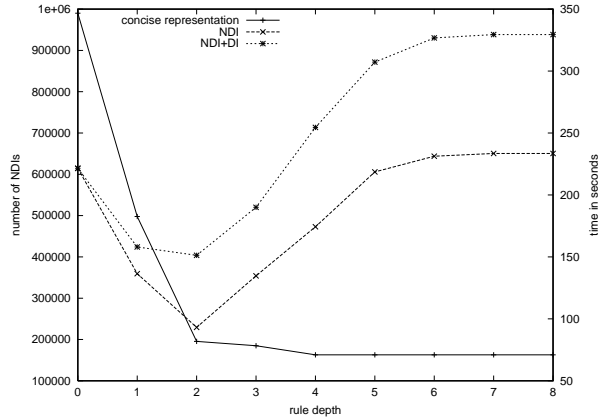,width=8cm} \caption{Strength of deduction
rules.}\label{fig:crsize}
\end{center}
\end{figure}

In the last experiment, we compared the strength of evaluating the
deduction rules up to a certain depth, and the time needed to
generate all NDIs w.r.t. the given depth.
Figure~\ref{fig:crsize} shows the results. On the x-axis, we show
the depth up to which rules are evaluated. We denoted the standard
Apriori monotonicity check by $0$, although it is actually
equivalent to the rules of depth $1$. The reason for this is that
we also used the other optimizations described in Section~3. More
specifically, if the lower or upper bound of an itemset equals its
actual support, we can prune its supersets, which is denoted as
depth $1$ in this figure. The left y-axis shows the number of NDIs
w.r.t. the given depth and is represented by the line `concise
representation'. The line `NDI' shows the time needed to generate
these NDIs. The time is shown on the right y-axis. The `NDI+DI'
line shows the time needed to generate all NDIs plus the time
needed to derive all DIs, resulting in all frequent patterns. As
can be seen, the size of the concise representation drops quickly
only using the rules of depth $1$ and $2$. From there on, higher
depths result in a slight decrease of the number of NDIs. From
depth $4$ on, this size stays the same, which is not that
remarkable since the number of NDIs of these sizes is also small.
The time needed to generate these sets is best if the rules are
only evaluated up to depth $2$. Still, the running time is almost
always better than the time needed to generate all frequent
itemsets (depth 0), and is hardly higher for higher depths. For
higher depths, the needed time increases, which is due to the
number of rules that need to be evaluated. Also note that the
total time required for generating all NDIs and deriving all DIs is
also better than generating all frequent patterns at once, at
depth $1$,$2$,and $3$.  This is due to the fact that the NDI
algorithm has to perform less scans through the transaction
database.  For larger databases this would also happen for the
other depths, since the derivation of all DIs requires no scan
through the database at all.

\section{Related Work}
\subsection{Concise Representations}
In the literature, there exist already a number of concise representations
for frequent itemsets. The most important ones are {\em closed itemsets\/}, {\em free itemsets\/},
and {\em disjunction-free itemsets\/}. We compare the different concise representations with the
NDI-representation.

\paragraph{Free sets \cite{BOULICAUT2000} or Generators \cite{Kryszkiewicz2001}} An itemset $I$ is
called {\em free\/} if it has no subset with the same support. We will denote the set of all
frequent free itemsets with $\mathit{FreqFree}$. In \cite{BOULICAUT2000}, the authors show that freeness
is anti-monotone; the subset of a free set must also be free. $\mathit{FreqFree}$ itself is not a concise
representation for the frequent sets, unless if the set 
$\mathit{Border}(\mathit{FreqFree}):=\{I\subseteq{\cal I}\mid \forall J\subset I: J\in \mathit{FreqFree}
 \wedge I \notin \mathit{FreqFree}\}$ is added~\cite{BOULICAUT2000}.
We call the concise representation consisting of these two sets $\mathit{ConFreqFree}$. Notice that free sets
\cite{BOULICAUT2000} and generators~\cite{PASQUIER99,Kryszkiewicz2001} are the same.

\paragraph{Disjunction-free sets \cite{BYKOWSKI2001} or disjunction-free generators~\cite{Kryszkiewicz2001}}
{\em Disjunction-free\/} sets are essentially an extension of free sets. A set $I$ is called
disjunction-free if there does not exist two items $i_1,i_2$ in $I$ such that
${\it support}(I)={\it support}(I-\{i_1\})+{\it support}(I-\{i_2\})-{\it support}(I-\{i_1,i_2\})$. This rule is
in fact our rule ${\cal R}_I(I-\{i_1,i_2\})$. Notice that free sets are a special case of this case,
namely when $i_1=i_2$. We will denote the set of frequent disjunction-free sets by $\mathit{FreqDFree}$.
Again, disjunction-freeness is anti-monotone, and $\mathit{FreqDFree}$ is not a concise representation of the
set of frequent itemsets, unless we add the border of $\mathit{FreqDFree}$. We call the concise representation
containing these two sets $\mathit{ConFreqDFree}$.

\paragraph{Closed itemsets \cite{PASQUIER99}} Another type of concise representation that received
a lot of attention in the literature \cite{BOULICAUT2000b,PEI2000b,ZAKI99} are the closed itemsets.
They can be introduced as follows: the {\em closure\/} of an itemset $I$ is the largest superset
of $I$ such that its support equals the support of $I$. This superset is unique and is denoted by
$cl(I)$. An itemset is called {\em closed\/} if it equals its closure. We will denote the set of all
frequent closed itemsets by $\mathit{FreqClosed}$. In \cite{PASQUIER99}, the authors show that $\mathit{FreqClosed}$
is a concise representation for the frequent itemsets.

In the following proposition we give connections between the different concise
representations.
\begin{proposition}
For every dataset and support threshold, the following inequalities are valid.
\begin{enumerate}
\item The set of frequent closed itemsets is always smaller or equal in cardinality than the set of frequent free sets. 
\item The set of NDIs is always a subset of $\mathit{ConFreqDFree}$.
\end{enumerate}
\end{proposition}
\begin{proof}
\begin{enumerate}
\item We first show that $\mathit{Closed}=\mathit{cl}(\mathit{Free})$.
\begin{itemize}
\item[$\subseteq$] Let $C$ be a closed set. Let $I$ be
a smallest subsets of $C$ such that $cl(I)=C$. Suppose $I$ is not a free set. Then there exist
$J\subset I$ such that ${\it support}(J)={\it support}(I)$. This rule however implies that ${\it support}(J)
={\it support}(C)$. This is in contradiction with the minimality of $I$.
\item[$\supseteq$] Trivial, since $cl$ is idempotent.
\end{itemize}
This equality implies that $\mathit{cl}$ is always a surjective function
from $\mathit{Free}$ to $\mathit{Closed}$, and therefore, $|\mathit{Free}|\geq |\mathit{Closed}|$.

\item Suppose $I$ is not in $\mathit{ConFreqDFree}$. If $I$ is not frequent, then the result is trivially satisfied.
Otherwise, this means that $I$ is not a frequent free set, and that there is at least one subset $J$
of $I$ that is also not a frequent free set (otherwise $I$ would be in the border of $\mathit{FreqDFree}$.)
Therefore, there exist $i_1,i_2\in J$ such that ${\it support}(J)={\it support}(J-\{i_1\})+{\it support}(J-\{i_2\})-
{\it support}(J-\{i_1,i_2\})=\sigma(J,J-\{i_1,i_2\})$. We now conclude, using Lemma~\ref{lem:lemma1}, that $I$ is a
derivable itemset, and thus not in NDI.
\end{enumerate}
\end{proof}
Other possible inclusions between the described concise representations
do not satisfy, i.e., for some datasets
and support thresholds we have $|\mathrm{NDI}|<|\mathit{Closed}|$, 
while other datasets and support thresholds have $|\mathit{Closed}|<|\mathrm{NDI}|$. 
We omit the proof of this due to space limitations. 
We should however mention that even
though $\mathit{FreqDFree}$ is always a superset of NDI, in the experiments
the gain of evaluating the extra rules is often small. In many
cases the reduction of $\mathit{ConFreqDFree}$, which corresponds to
evaluating rules up to depth 2 in our framework, is almost as big
as the reduction using the whole set of rules. Since our rules are
complete, this shows that additional gain is in many cases
unlikely.

\subsection{Counting Inference}
\paragraph{MAXMINER \cite{BAYARDO98}}
In MAXMINER, {\it Bayardo\/} uses the following rule to derive a lower bound on the support of
an itemset: $${\it support}(I\cup\{i\})\leq {\it support}(I)-\sum_{j\in T}{\it drop}(J,j)$$
with $T=I-J$, $J\subset I$, and ${\it drop}(J,j)={\it support}(J)-{\it support}(J\cup\{j\})$.
This derivation corresponds to repeated application of rules ${\cal R}_I(I-\{i_1,i_2\})$.

\paragraph{PASCAL \cite{BASTIDE2000}}
In their PASCAL-algorithm, {\it Bastide et al.} use counting inference to avoid counting the
support of all candidates. The rule they are using to avoid counting is based on our rule
${\cal R}_I(I-\{i\})$. In fact the PASCAL-algorithm corresponds to our
algorithm when we only check rules up to depth 1, and do not prune derivable sets. Instead of
counting the derivable sets, we use the derived support. Here the same remark as with the
$\mathit{ConFreqDFree}$-representation applies; although PASCAL does not use all rules, in many cases the
performance comes very close to evaluating all rules, showing that for these databases PASCAL is
nearly optimal.

\bibliographystyle{plain}

\begin{thebibliography}{10}

\bibitem{AGRAWAL93}
R.~Agrawal, T.~Imilienski, and A.~Swami.
\newblock Mining association rules between sets of items in large databases.
\newblock In {\em Proc. ACM SIGMOD Int. Conf. Management of Data}, pages
  207--216, Washington, D.C., 1993.

\bibitem{AGRAWAL94}
R.~Agrawal and R.~Srikant.
\newblock Fast algorithms for mining association rules.
\newblock In {\em Proc. VLDB Int. Conf. Very Large Data Bases}, pages 487--499,
  Santiago, Chile, 1994.

\bibitem{BASTIDE2000}
Y.~Bastide, R.~Taouil, N.~Pasquier, G.~Stumme, and L.~Lakhal.
\newblock Mining frequent patterns with counting inference.
\newblock {\em ACM SIGKDD Explorations}, 2(2):66--74, 2000.

\bibitem{BAYARDO98}
R.~J. Bayardo.
\newblock Efficiently mining long patterns from databases.
\newblock In {\em Proc. ACM SIGMOD Int. Conf. Management of Data}, pages
  85--93, Seattle, Washington, 1998.

\bibitem{BOULICAUT2000b}
J.-F. Boulicaut and A.~Bykowski.
\newblock Frequent closures as a concise representation for binary data mining.
\newblock In {\em Proc. PaKDD Pacific-Asia Conf. on Knowledge Discovery and
  Data Mining}, pages 62--73, 2000.

\bibitem{BOULICAUT2000}
J.-F. Boulicaut, A.~Bykowski, and C.~Rigotti.
\newblock Approximation of frequency queries by means of free-sets.
\newblock In {\em Proc. PKDD Int. Conf. Principles of Data Mining and Knowledge
  Discovery}, pages 75--85, 2000.

\bibitem{BYKOWSKI2001}
A.~Bykowski and C.~Rigotti.
\newblock A condensed representation to find frequent patterns.
\newblock In {\em Proc. PODS Int. Conf. Principles of Database Systems}, 2001.

\bibitem{CALDERSEDBT2002}
T.~Calders.
\newblock Deducing bounds on the frequency of itemsets.
\newblock In {\em EDBT Workshop DTDM Database Techniques in Data Mining}, 2002.

\bibitem{UCIKDD}
S.~Hettich and S.~D. Bay.
\newblock {\em The UCI KDD Archive. [http://kdd.ics.uci.edu]}.
\newblock Irvine, CA: University of California, Department of Information and
  Computer Science, 1999.

\bibitem{KNUTHIE97}
D.E. Knuth.
\newblock {\em Fundamental Algorithms}.
\newblock Addison-Wesley, Reading, Massachusetts, 1997.

\bibitem{Kryszkiewicz2001}
M.~Kryszkiewicz.
\newblock Concise representation of frequent patterns based on disjunction-free
  generators.
\newblock In {\em Proc. IEEE Int. Conf. on Data Mining}, pages 305--312, 2001.

\bibitem{MANNILA96}
H.~Mannila and H.~Toivonen.
\newblock Multiple uses of frequent sets and condensed representations.
\newblock In {\em Proc. KDD Int. Conf. Knowledge Discovery in Databases}, 1996.

\bibitem{PASQUIER99}
N.~Pasquier, Y.~Bastide, R.~Taouil, and L.~Lakhal.
\newblock Discovering frequent closed itemsets for association rules.
\newblock In {\em Proc. ICDT Int. Conf. Database Theory}, pages 398--416, 1999.

\bibitem{PEI2000b}
J.~Pei, J.~Han, and R.~Mao.
\newblock Closet: An efficient algorithm for mining frequent closed itemsets.
\newblock In W.~Chen, J.F. Naughton, and P.A. Bernstein, editors, {\em ACM
  SIGMOD Workshop on Research Issues in Data Mining and Knowledge Discovery},
  Dallas, TX, 2000.

\bibitem{ZAKI99}
M.J. Zaki and C.~Hsiao.
\newblock {ChARM}: An efficient algorithm for closed association rule mining.
\newblock In {\em Technical Report 99-10, Computer Science, Rensselaer
  Polytechnic Institute}, 1999.

\bibitem{realworld}
Z.~Zheng, R.~Kohavi, and L.~Mason.
\newblock Real world performance of association rule algorithms.
\newblock In {\em Proc. KDD Int. Conf. Knowledge Discovery in Databases}, pages
  401--406. ACM Press, 2001.

\end{thebibliography}

\end{document}